\documentclass[prl,twocolumn,showpacs]{revtex4}
\usepackage{amsfonts}
\usepackage{amsmath}
\usepackage{dcolumn}
\usepackage{graphicx}
\usepackage{bm}

\begin{document}

\title{Projected BCS states and spin Hamiltonians for the SO($n$)$_{1}$ WZW
model}
\author{Hong-Hao Tu}
\affiliation{Max-Planck-Institut f\"ur Quantenoptik, Hans-Kopfermann-Str. 1, D-85748
Garching, Germany}
\date{\today}

\begin{abstract}
We propose a class of projected BCS wave functions and derive their parent
spin Hamiltonians. These wave functions can be formulated as infinite Matrix
Product States constructed by chiral correlators of Majorana fermions. In
1D, the spin Hamiltonians can be viewed as SO($n$) generalizations of
Haldane-Shastry models. We numerically compute the spin-spin correlation
functions and R\'{e}nyi entropies for $n=5$ and $6$. Together with the
results for $n=3$ and $4$, we conclude that these states are critical and
their low-energy effective theory is the SO($n$)$_{1}$ Wess-Zumino-Witten
model. In 2D, we show that the projected BCS states are chiral spin liquids,
which support non-Abelian anyons for odd $n$ and Abelian anyons for even $n$.
\end{abstract}

\maketitle

\textit{Introduction.-- }An efficient description of quantum many-body systems
is a challenging problem in modern physics, as the dimension of the Hilbert
space scales exponentially with the number of particles. For strongly
interacting many-body systems, much of our understanding of their properties
comes from physically motivated trial wave functions and/or exact solutions
of specific models. A great success of the trial wave function approach is
the celebrated Laughlin wave function for the\ fractional quantum Hall
effect at $1/m$ (with $m$ odd) filling \cite{Laughlin-1983}. Toward exact
results, Bethe's solution of the spin-1/2 Heisenberg chain \cite{Bethe-1931}\
and integrability of the spin-1/2 Haldane-Shastry model \cite%
{Haldane-Shastry-1988} provide invaluable insight for critical spin-1/2
chains.

The justification of trial wave functions is usually a difficult task. For
example, the relevance of Anderson's resonating valence bond (RVB) state
\cite{Anderson-1987}\ for the mechanism of high-$T_{c}$ superconductivity is
still a controversial issue. A useful technique for justifying trial wave
functions is to study their parent Hamiltonians for which the trial wave
functions are exact ground states. For the Laughlin wave function, the parent
Hamiltonian which consists of certain Haldane pseudopotentials \cite%
{Haldane-1983} differs from physical Coulomb interactions but their
difference can be viewed as a perturbation \cite{Haldane-Rezayi-1985}.
A similar situation arises for the spin-1 Affleck-Kennedy-Lieb-Tasaki (AKLT) state and its parent
Hamiltonian \cite{Affleck-1987}, which contains an extra biquadratic term
apart from Heisenberg interactions. Since the spin-1 AKLT model can be
adiabatically connected to the spin-1 Heisenberg chain without closing the
gap, it is widely believed that the AKLT state qualitatively captures the
physics of the spin-1 Heisenberg chain.

In this work, we propose a class of projected BCS states and derive their
parent Hamiltonians. These states can also be represented as infinite Matrix
Product States (MPSs) \cite{Ignacio-German-2010} constructed from chiral
correlators of Majorana fermions. In 1D, the spin Hamiltonians are SO($n$)
generalizations of Haldane-Shastry models. We numerically calculate the
spin-spin correlation functions and the R\'{e}nyi entropies for $n=5$ and $6$
and compare the numerical results with field theory predictions from SO($n$)$%
_{1}$ criticality. Together with the known results for $n=3$ and $4$, we
expect that for general $n$ these states are critical and belong to the SO($%
n $)$_{1}$ Wess-Zumino-Witten (WZW) universality class. We also show that the projected BCS
states with modified Cooper pair wave functions provide a good description
for Ising ordered and disordered phases close to SO($n$)$_{1}$ criticality.
In 2D, the projected BCS states are chiral spin liquids with $p+ip$ pairing
symmetry. We find that these topological states support non-Abelian Ising
anyons for odd $n$ and Abelian anyons for even $n$, respectively.

\begin{figure}[tbp]
\centering
\includegraphics[scale=0.4]{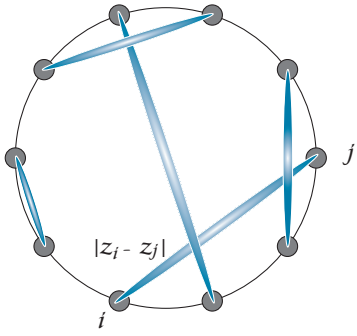}
\caption{(Color online) Schematic of a valence-bond configuration in the
projected BCS state (\protect\ref{eq:SO(n)HS}). The valence bonds (blue) are
SO($n$) singles formed by two SO($n$) vectors. In the uniform case $%
z_{j}=\exp (i\frac{2\protect\pi }{N}j)$, the periodic chain is viewed as a
unit circle and $|z_{i}-z_{j}|$ is the chord distance between two sites. }
\label{fig:rvb}
\end{figure}

\textit{Projected BCS wave function.-- }Constructing the projected BCS wave
functions relies on a slave-particle representation of the SO($n$) algebra.
Let us start from a 1D periodic chain with even $N$ sites, where the $n$
vectors in each site are represented by using singly occupied fermions, $%
|n^{a}\rangle =c_{a}^{\dagger }|0\rangle $ ($a=1,\ldots ,n$). In terms of
fermions, the SO($n$) generators are written as $L^{ab}=i(c_{a}^{\dagger
}c_{b}-c_{b}^{\dagger }c_{a})$, where $1\leq a<b\leq n$. To remove
unphysical states in this fermionic representation, a single-occupancy
constraint is required, $\sum_{a=1}^{n}c_{j,a}^{\dagger }c_{j,a}=1$ $\forall
j=1,\ldots ,N$, which defines a Gutzwiller projector $P_{\mathrm{G}}$. Then,
the projected BCS wave function of our interest is defined by
\begin{equation}
|\Psi \rangle =P_{\mathrm{G}}\exp \left( \sum_{i<j}\frac{1}{z_{i}-z_{j}}%
\sum_{a=1}^{n}c_{i,a}^{\dagger }c_{j,a}^{\dagger }\right) |0\rangle ,
\label{eq:SO(n)HS}
\end{equation}%
where $\sum_{a=1}^{n}c_{i,a}^{\dagger }c_{j,a}^{\dagger }$ creates an SO($n$%
) singlet between sites $i$ and $j$. Note that $|\Psi \rangle $ is a
coherent superposition of valence-bond singlets of arbitrary range (see Fig. %
\ref{fig:rvb}) and hence can be viewed as an RVB state \cite{Anderson-1987}%
. If we choose $z_{j}=\exp (i\frac{2\pi }{N}j)$, the amplitude of the Cooper
pair wave function $1/|z_{i}-z_{j}|$ is the inverse of the chord distance
between the sites. Under this choice, $|\Psi \rangle $ is both real and
translationally invariant, which is the uniform case that we will consider
in the following.

Before discussing the properties of $|\Psi \rangle $ for general $n$, we
establish the relation between (\ref{eq:SO(n)HS}) and some known results.
For $n=3$, after switching to the standard spin-1 basis $|n^{1}\rangle =%
\frac{1}{\sqrt{2}}(|-1\rangle -|1\rangle )$, $|n^{2}\rangle =\frac{i}{\sqrt{2%
}}(|-1\rangle +|1\rangle )$, $|n^{3}\rangle =|0\rangle $, we find that the
projected BCS state (\ref{eq:SO(n)HS}) is equivalent to the spin-1
Haldane-Shastry state, which has been considered in Refs. \cite%
{Anne-2011,Greiter-2012,Paredes-2012}. It was shown \cite%
{Anne-2011,Greiter-2012} that this state is critical and its low-energy
effective theory is an SU(2)$_{2}$ (or equivalently SO(3)$_{1}$) WZW model.
In a recent work \cite{ZXLiu-2012}, a projected BCS wave function similar to
(\ref{eq:SO(n)HS}) is used as a variational ansatz to describe the phases in the
spin-1 bilinear-biquadratic chain, including the Takhtajan-Babujian model
\cite{Takhtajan-Babujian-1982} which also belongs to the SU(2)$_{2}$ WZW
universality class \cite{Affleck-1986}. For $n=4$, after representing the
four SO(4) vectors by two spin-1/2 states, we find that the projected BCS
state (\ref{eq:SO(n)HS}) can be rewritten as a product of two decoupled
spin-1/2 Haldane-Shastry states. An immediate consequence of this
decomposition is that the SO(4) state is critical and represents the fixed
point of the SO(4)$_{1}$ WZW model.

\textit{Infinite MPS and parent Hamiltonian.--} From the known results for $%
n=3$ and $4$, one may speculate that for general $n$ the projected BCS state
(\ref{eq:SO(n)HS}) belongs to the SO($n$)$_{1}$ WZW universality class. Let
us further uncover this relationship by formulating (\ref{eq:SO(n)HS}) as an
infinite MPS. The SO($n$)$_{1}$ WZW model has a primary field with conformal
weight $h_{v}=1/2$ in the vector representation \cite{YellowBook}, which can
be naturally interpreted as Majorana fermion fields $\chi ^{a}(z)$ ($%
a=1,\ldots ,n$). This Majorana representation of the primary field allows us
to rewrite the projected BCS state (\ref{eq:SO(n)HS}) as the following
infinite MPS:
\begin{equation}
|\Psi \rangle =\sum_{a_{1},\ldots ,a_{N}=1}^{n}\Psi (a_{1},\ldots
,a_{N})|n^{a_{1}},\ldots ,n^{a_{N}}\rangle ,  \label{eq:iMPS}
\end{equation}%
where the coefficients are the chiral correlators of $N$ Majorana fields
\cite{Ardonne-German-2010}%
\begin{equation}
\Psi (a_{1},\ldots ,a_{N})=\langle \chi ^{a_{1}}(z_{1})\chi
^{a_{2}}(z_{2})\cdots \chi ^{a_{N}}(z_{N})\rangle .  \label{eq:iMPS1}
\end{equation}%
A detailed proof of the equivalence of the projected BCS state (\ref%
{eq:SO(n)HS}) and the infinite MPS (\ref{eq:iMPS1}) is given in the
Supplemental Material \cite{SuppleMatSOn}.

Unlike usual MPS with finite matrix dimensions, the state (\ref{eq:iMPS1})
is an infinite MPS \cite{Ignacio-German-2010}\ since its ancillary space on
which the Majorana fields act is an infinite-dimensional Hilbert space. This
allows the infinite MPS (\ref{eq:iMPS1}) to describe the expected SO($n$)$%
_{1}$ criticality with unbounded increase of the entanglement entropy.

The key benefit of the infinite MPS formulation is that a parent Hamiltonian
can be derived, such that (\ref{eq:iMPS1}) is its exact ground state. As
shown in Ref. \cite{Anne-2011}, the presence of null vectors in conformal
field theories (CFTs) leads to a set of operators which annihilate the
infinite MPS. Following this approach, we have derived \cite{SuppleMatSOn}
such\ operators for (\ref{eq:iMPS1})
\begin{equation*}
\Lambda _{i}^{ab}=\sum_{j(\neq i)}\frac{w_{ij}}{3}[2L_{j}^{ab}-\frac{3}{n-1}%
L_{i}^{ab}(\vec{L}_{i}\cdot \vec{L}_{j})+(\vec{L}_{i}\cdot \vec{L}%
_{j})L_{i}^{ab}],
\end{equation*}%
where $w_{ij}\equiv (z_{i}+z_{j})/(z_{i}-z_{j})$ and $\vec{L}_{i}\cdot \vec{L%
}_{j}\equiv \sum_{a<b}L_{i}^{ab}L_{j}^{ab}$. Since $\Lambda _{i}^{ab}|\Psi
\rangle =0$ $\forall i,a,b$ and $\sum_{i}L_{i}^{ab}|\Psi \rangle =0$ $%
\forall a,b$, we can define a parent Hamiltonian $H=\sum_{i,a<b}(\Lambda
_{i}^{ab})^{\dagger }\Lambda _{i}^{ab}+\frac{2(N-2)}{3}\sum_{a<b}(%
\sum_{i}L_{i}^{ab})^{2}+E_{0}$ whose ground state is the infinite MPS (\ref%
{eq:iMPS1}) with energy $E_{0}$. Choosing $E_{0}=-\frac{2}{9}(n-1)N(N^{2}-4)$%
, the explicit form of $H$ is given by%
\begin{eqnarray}
H &=&-\sum_{i\neq j}w_{ij}^{2}[\frac{n+2}{3}(\vec{L}_{i}\cdot \vec{L}_{j})+%
\frac{n-4}{3(n-1)}(\vec{L}_{i}\cdot \vec{L}_{j})^{2}]  \notag \\
&&-\frac{n-4}{3(n-1)}\sum_{i\neq j\neq k}w_{ij}w_{ik}(\vec{L}_{i}\cdot \vec{L%
}_{j})(\vec{L}_{i}\cdot \vec{L}_{k}).  \label{eq:Hamiltonian}
\end{eqnarray}%
Generically, the Hamiltonian (\ref{eq:Hamiltonian}) is a long-ranged SO($n$%
)\ bilinear-biquadratic model with three-spin interactions. For $n=4$, as we
expected, the Hamiltonian only has inverse-square Heisenberg exchange
interactions, which can be decomposed into two spin-1/2 Haldane-Shastry
Hamiltonians due to SO(4)$\simeq $SU(2)$\times $SU(2).

\textit{Jastrow versus Pfaffian.--} It is well known that the SO($n$) Lie
algebra has a sharp difference between even and odd $n$ \cite{Georgi-1999}.
As we shall see, this leads to distinct forms of the wave function (\ref%
{eq:SO(n)HS}) in the Cartan basis for even and odd $n$: The former has a pure
Jastrow form, while the latter includes a Pfaffian factor. To see this
difference, let us consider SO($2l$) and SO($2l+1$) ($l$: integer) and
choose mutually commuting Cartan generators as $L^{12},L^{34},\ldots
L^{2l-1,2l}$. For SO($2l$), a convenient choice of the Cartan basis is
defined by $|0,\ldots ,m_{\alpha }=\pm 1,\ldots ,0\rangle =(|n^{2\alpha
}\rangle \pm i|n^{2\alpha -1}\rangle )/\sqrt{2}$ ($\alpha =1,\ldots ,l$),
where $m_{\alpha }$ is the eigenvalue of $L^{2\alpha -1,2\alpha }$. For the
vectors $|0,\ldots ,m_{\alpha }=\pm 1,\ldots ,0\rangle $, we label their
positions in the spin chain by $x_{1}^{(\alpha )}<\cdots <x_{N_{\alpha
}}^{(\alpha )}$. In this basis, the wave function (\ref{eq:SO(n)HS}) for
even $n=2l$ takes the form \cite{SuppleMatSOn}%
\begin{equation}
\Psi (\{m\})=\rho _{m}\prod_{\alpha
=1}^{l}\prod_{i<j}(z_{i}-z_{j})^{m_{\alpha ,i}m_{\alpha ,j}},
\label{eq:evenSOn}
\end{equation}%
where $\rho _{m}=\mathrm{sgn}(x_{1}^{(1)}\ldots x_{N_{1}}^{(1)}\ldots
x_{1}^{(l)}\ldots x_{N_{l}}^{(l)})$ ($\mathrm{sgn}$: signature of a
permutation) if $\sum_{i}m_{\alpha ,i}=0$ $\forall \alpha $ and $\rho _{m}=0$
otherwise.

For SO($2l+1$), apart from the vectors $|0,\ldots ,m_{\alpha }=\pm 1,\ldots
,0\rangle $, there exists an additional vector $|0,\ldots ,0\rangle
=|n^{2l+1}\rangle $, which is annihilated by all Cartan generators. Labeling
their positions by $x_{1}^{(0)}<\cdots <x_{N_{0}}^{(0)}$, the wave function (\ref{eq:SO(n)HS}) for odd $n=2l+1$ is written as \cite{SuppleMatSOn}%
\begin{equation}
\Psi (\{m\})=\rho _{m}\mathrm{Pf}_{0}(\frac{1}{z_{i}-z_{j}})\prod_{\alpha
=1}^{l}\prod_{i<j}(z_{i}-z_{j})^{m_{\alpha ,i}m_{\alpha ,j}},
\label{eq:oddSOn}
\end{equation}%
where $\rho _{m}=\mathrm{sgn}(x_{1}^{(0)}\ldots
x_{N_{0}}^{(0)},x_{1}^{(1)}\ldots x_{N_{1}}^{(1)}\ldots x_{1}^{(l)}\ldots
x_{N_{l}}^{(l)})$ if $\sum_{i}m_{\alpha ,i}=0$ $\forall \alpha $ and $\rho
_{m}=0$ otherwise, and the Pfaffian factor $\mathrm{Pf}_{0}(\frac{1}{%
z_{i}-z_{j}})$ is restricted to the positions for the extra vector $%
|0,\ldots ,0\rangle $.

\textit{Numerical results.-- }The power-law decaying correlation functions
and the universal scaling of entanglement entropy \cite{Cardy-Calabrese-2004}
are characteristic behaviors of conformal critical points in 1D. Even though
these quantities are difficult to compute analytically for (\ref%
{eq:SO(n)HS}), the Jastrow and Pfaffian forms (\ref{eq:evenSOn}) and (\ref%
{eq:oddSOn}) of the wave functions are very suitable for determining them
numerically via the Metropolis Monte Carlo (MC) method \cite%
{Horsch-Kaplan-1983}. Below we focus on the projected BCS state (\ref%
{eq:SO(n)HS}) with $n=5$ and $6$ and provide numerical evidence that they
belong to SO(5)$_{1}$\ and SO(6)$_{1}$ WZW models, respectively.

\begin{figure}[tbp]
\centering
\includegraphics[scale=0.5]{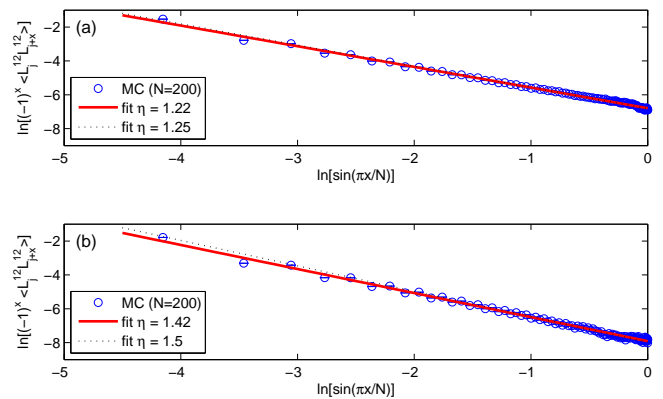}
\caption{(Color online) Spin-spin correlation function (logarithmic scale)$%
\ \ln [(-1)^{x}\langle L_{j}^{12}L_{j+x}^{12}\rangle ]$\ as a function of $%
\ln [\sin (\protect\pi x/N)]$ in the projected BCS state (\protect\ref%
{eq:SO(n)HS}) for $N=200$ and (a) $n=5$ and (b) $n=6$. The solid lines (red)
are fits of the form $\ln [(-1)^{x}\langle L_{j}^{12}L_{j+x}^{12}\rangle ]=%
\protect\eta \ln [\sin (\protect\pi x/N)]+A$, where $\protect\eta $ and $A$
are fitting parameters. The dotted lines are also fits of this formula but
with the field theory prediction $\protect\eta =n/4$ (Ref. \protect\cite%
{Tu-2011}).}
\label{fig:correlator}
\end{figure}

For critical spin chains in the SO($n$)$_{1}$ WZW universality class, field
theory predicts that for $n<8$ the spin-spin correlation function behaves as
$\langle L_{j}^{ab}L_{j+x}^{ab}\rangle \sim (-1)^{x}/x^{\eta }$ with $\eta
=n/4$ \cite{Tu-2011}. For the projected BCS state (\ref{eq:SO(n)HS}) with $%
n=5$ and $6$, we have computed the two-point spin correlator $\langle
L_{j}^{12}L_{j+x}^{12}\rangle $. Figure \ref{fig:correlator} shows the
numerical results for $N=200$. The critical exponents that best fit with our
numerical data are $\eta =1.22$ for SO(5) and $\eta =1.42$ for SO(6) (solid
lines in Fig. \ref{fig:correlator}), which agree very well with the field
theory predictions (dotted lines).

\begin{figure}[tbp]
\centering
\includegraphics[scale=0.5]{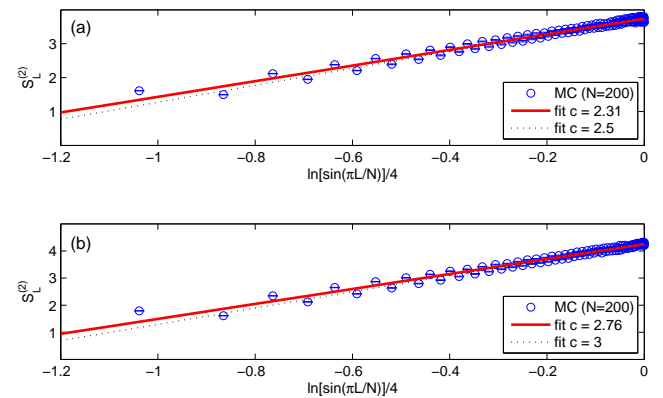}
\caption{(Color online) R\'{e}nyi entropy $S_{L}^{(2)}$\ as a function of $%
\ln [\sin (\protect\pi L/N)]/4$ in the projected BCS state (\protect\ref%
{eq:SO(n)HS}) for $N=200$ and (a) $n=5$ and (b) $n=6$. The solid lines (red)
are fits of the form $S_{L}^{(2)}=c\ln [\sin (\protect\pi %
L/N)]/4+c_{2}^{\prime }$, where $c$ and $c_{2}^{\prime }$ are fitting
parameters. The dotted lines are also fits of this formula but the central
charge $c$ is fixed to $c=n/2$ of the SO($n$)$_{1}$ WZW model.}
\label{fig:entropy}
\end{figure}

The entanglement entropy that is easily accessible via the MC method is the R%
\'{e}nyi entropy $S_{L}^{(2)}=-\ln \mathrm{Tr}\rho _{L}^{2}$ (see Refs. \cite
{Ignacio-German-2010,Hastings-2010,YZhang-2011,Fabio-2012}), where $\rho
_{L} $ is the reduced density matrix of the state in a subsystem of length $%
L $. For the SO($n$)$_{1}$ WZW model with $c=n/2$ we expect $S_{L}^{(2)}=c\ln
[\sin (\pi L/N)]/4+c_{2}^{\prime }$ \cite{Cardy-Calabrese-2004}, where $%
c_{2}^{\prime }$ is a constant. For $n=5$ and $6$, we plot $S_{L}^{(2)}$ as
a function of $\ln [\sin (\pi L/N)]/4$ for $N=200$ in Fig. \ref{fig:entropy}. From our MC data, the estimates of the central charge are $c=2.31$ for
SO(5) and $c=2.76$ for SO(6) (solid lines in Fig. \ref{fig:entropy}), which
are close to the predicted $c=n/2$ (dotted lines) but show some deviations.

The origin of the small deviations of the numerical results and the SO($n$)$%
_{1}$ predictions may be due to the presence of marginally irrelevant terms
in the SO($n$)$_{1}$ WZW model for (\ref{eq:SO(n)HS}) and its parent
Hamiltonian (\ref{eq:Hamiltonian}), unlike the SU($n$) Haldane-Shastry
models \cite{Kawakami-1992,Ha-Haldane-1992} (including the spin-1/2
Haldane-Shastry model for $n=2$) being the fixed points of the SU($n$)$_{1}$
WZW model. For $n=3$, the presence of marginal term in the spin-1
Haldane-Shastry model has been confirmed numerically \cite%
{Anne-2011,Greiter-2012}. If this is also the case for $n\geq 5$, an
interesting open question is whether there exist a modified version of (\ref%
{eq:SO(n)HS}) and its parent Hamiltonian that represent the fixed point of
the SO($n$)$_{1}$ WZW model.

\textit{Away from SO(}$n$\textit{)}$_{1}$\textit{\ criticality.-- }After
showing that the projected BCS state (\ref{eq:SO(n)HS}) captures the physics
of the SO($n$)$_{1}$ WZW model, it is natural to ask whether similar
projected wave functions are relevant for gapped spin chains away from (but
close to) SO($n$)$_{1}$ criticality. Let us restrict ourselves to SO($n$)
symmetric models for simplicity. According to the well-known result by
Witten \cite{Witten-1984}, the SO($n$)$_{1}$ WZW model is equivalent to $n$
massless Majorana fermions, i.e., $n$ Ising models at criticality. For this
critical theory, the only relevant perturbation allowed by SO($n$) symmetry
is the mass term of Majorana fermions. Thus, the low-energy effective theory
has the Hamiltonian density $\mathcal{H}=-\frac{iv}{2}\sum_{\nu =1}^{n}(\xi
_{R}^{\nu }\partial _{x}\xi _{R}^{\nu }-\xi _{L}^{\nu }\partial _{x}\xi
_{L}^{\nu })-im\sum_{\nu =1}^{n}\xi _{R}^{\nu }\xi _{L}^{\nu }$, where $\xi
_{R(L)}^{\nu }$ are right (left) moving Majorana fermions, $v$ and $m$ are
their velocity and mass. Here we have assumed four-fermion interactions are
weak and can be neglected, since they are marginal and only renormalize the
mass of Majorana fermions at low-energy limit \cite{Gogolin-1998}.

The SO($n$)$_{1}$ criticality corresponds to $m=0$. The two gapped phases
adjacent to the SO($n$)$_{1}$ criticality are (i) the Ising ordered phase ($m<0$)
and (ii) the Ising disordered phase ($m>0$). For these two phases, we note that
they can be well described by modified projected BCS states. Actually, these
two gapped phases and an SO($n$)$_{1}$ critical point (Reshetikhin model
\cite{Reshetikhin-1985})\ are realized in the SO($n$) bilinear-biquadratic
chain \cite{Tu-2011,Tsvelik-Nersesyan-Schoutens-Lecheminant}.\ The ideal
example that belongs to the Ising ordered phase is the SO($n$) AKLT model
\cite{Affleck-1987,Tu-2008,Kolezhuk-Scalapino-Affleck}, whose ground state
can be represented as a projected BCS state, by replacing $%
g_{ij}=1/(z_{i}-z_{j})$ in (\ref{eq:SO(n)HS}) with $g_{ij}=1$ \cite%
{ZXLiu-2012}. For the Ising disordered phase, the ground state of the spin
chain is dimerized \cite{Tu-2011} and hence the valence bonds are
short ranged. In this case, a proper Cooper pair wave function for the
projected BCS state can be chosen as $g_{ij}\sim \exp (-|z_{i}-z_{j}|/\xi )$%
, where $\xi $ is the length scale of the valence bonds. In the extreme
case, a Cooper pair wave function that is nonvanishing only between
neighboring sites yields a Majumdar-Ghosh-like state, corresponding to perfect
dimerization. These results imply that both Ising ordered and disordered
phases close to SO($n$)$_{1}$ criticality are well described by projected
BCS states with properly chosen $g_{ij}$. Indeed, for $n=3$, it was shown
\cite{ZXLiu-2012}\ that the projected BCS states with Cooper pair wave
functions generated from Kitaev's Majorana chains \cite{Kitaev-2001} are
good variational wave functions for the Haldane (Ising ordered) and the
dimerized (Ising disordered) phases.

\textit{2D chiral spin liquids.-- }After establishing the relevance of
projected BCS states (\ref{eq:SO(n)HS}) for SO($n$)$_{1}$ criticality in 1D,
we move on and discuss their properties in a 2D square lattice, where the $z$%
's in (\ref{eq:iMPS1}) are now complex coordinates of the lattice sites. In
an analogy with fractional quantum Hall (FQH) states constructed by
conformal blocks of their gapless edge CFTs \cite{Moore-Read-1991,Anne-2012}%
, the chiral correlator (\ref{eq:iMPS1}) from the SO($n$)$_{1}$ WZW model ($n$
massless Majorana fermions) yields chiral spin liquids, which break time
reversal symmetry and are spin counterparts of FQH states \cite{XGWen-1989}.
From the projected BCS form (\ref{eq:SO(n)HS}), the Cooper pair wave
function $1/(z_{i}-z_{j})$ now corresponds to the topological phase of $p+ip$
superconductors \cite{Read-Green-2000} supporting chiral gapless Majorana
edge modes, which justifies the above bulk-edge correspondence. Below we
focus on the anyonic quasiparticle excitations in these 2D states, which
have intriguing properties depending on $n$ mod $16$, i.e., a 16-fold
way.

For odd $n$, the quasiparticles built upon the SO($n$) states support
non-Abelian statistics. Let us adapt the CFT approach of creating quasihole
excitations in FQH states \cite{Moore-Read-1991} to our spin system. For odd
$n$, the SO($n$)$_{1}$ WZW model has three primary fields: identity field\ $%
I $, vector field $v$, and spinor field $s$. Following the CFT approach,
creating quasiparticles in the SO($n$) state is achieved by adding spinor
fields $s$ in the chiral correlator (\ref{eq:iMPS1}). Then, the statistics
of quasiparticles are encoded in the fusion rules of the primary fields. In
fact, the spinor fields have a nontrivial fusion rule $s\times s=I+v$,
together with $s\times v=s$ and $v\times v=I$. These fusion rules resemble
those in Ising CFT ($\sigma \times \sigma =I+\varepsilon $, $\sigma \times
\varepsilon =\sigma $, and $\varepsilon \times \varepsilon =I$), which are
responsible for the non-Abelian statistics of Ising anyons \cite{Kitaev-2006}%
. Indeed, the Majorana free field representation of SO($n$)$_{1}$ WZW model
allows us to identify the spinor fields $s$ with conformal weight $%
h_{s}=n/16 $ as a product of $n$ Ising $\sigma $ fields ($h_{\sigma }=1/16$%
). Thus, we conclude that the SO($n$) states support non-Abelian Ising
anyons for odd $n$. Note that the case with $n=3$ recovers the physics of
the Moore-Read states \cite{Moore-Read-1991,Greiter-2009}, while for odd $%
n\geq 5$ they are natural generalizations of the Moore-Read states.

Now we show that the SO($n$) states only support Abelian anyons for even $n$%
. This subtle difference roots in the fusion rules of the SO($n$)$_{1}$
primary fields. In contrast to odd $n$ case, the SO($n$)$_{1}$ WZW model
with even $n$ has two spinor primary fields $s_{+}$ and $s_{-}$ with
conformal weight $h_{s_{+}}=h_{s_{-}}=n/16$ \cite{YellowBook}, apart from
the usual identity and vector fields. The fusion rules of spinor and vector
fields are $s_{+}\times v=s_{-}$ and $s_{-}\times v=s_{+}$. Depending on the
parity of $n/2$, the fusion rules involving two spinor fields are $%
s_{+}\times s_{+}=s_{-}\times s_{-}=I$, $s_{+}\times s_{-}=v$ for even $n/2$
and $s_{+}\times s_{+}=s_{-}\times s_{-}=v$, $s_{+}\times s_{-}=I$ for odd $%
n/2$ \cite{Schoutens-1999}. However, due to the absence of multiplicity in
the fusion outcome, only Abelian anyons can exist in the SO($n$) states with
even $n$.

More precisely, the anyonic properties of the SO($n$) states depend on $n$
mod $16$ (16-fold way) \cite{Sixteenfoldway}. The topological spin of
the quasiparticles generated by SO($n$)$\ $spinor primary fields is $\theta
_{s}=e^{i2\pi h_{s}}=e^{in\pi /8}$ (for both odd and even $n$), which is a
clear signature of the 16-fold way. For example, the quasiparticles $%
s_{+}$ and $s_{-}$ for SO(8) have $\theta _{s_{+}}=\theta _{s_{-}}=$ $-1$,
which are both fermions. Actually, this 16-fold way of the anyonic
properties has been analyzed in detail by Kitaev. In Ref. \cite{Kitaev-2006}%
, he considered a theory with $Z_{2}$ vortices and\ free Majorana fermions
whose energy band has Chern number $\nu $ and showed that the anyonic
properties of the unpaired Majorana modes in the vortex core depends on $\nu
$ mod $16$. Thus, our present work shows that the SO($n$)$_{1}$ CFT is
responsible for this 16-fold way and provides a class of microscopic
Hamiltonians which realize this interesting physics.

\textit{Conclusion and perspective.--} To conclude, we have proposed a class
of projected BCS states and derived their parent Hamiltonians. These states
also have an infinite MPS form generated by chiral correlators of Majorana
fermions. In 1D, they can be viewed as SO($n$) generalizations of
Haldane-Shastry models and capture the physics of the SO($n$)$_{1}$ WZW model.
These results indicate that modified projected BCS states are good
variational ansatz for describing Ising ordered and disordered phases close
to SO($n$)$_{1}$ criticality. In 2D, the SO($n$) states are chiral spin
liquid states, which support non-Abelian Ising anyons for odd $n$ and
Abelian anyons for even $n$. An open question that deserves further
investigation is whether these 2D chiral spin liquids are relevant for
physical models and materials \cite{TKNg-PALee-2010}. Moreover, our 2D toy
models may also shed light on another interesting open question: Can $p+ip$
pairing states arise after doping these antiferromagnets?

The author is indebted to G. Sierra for numerous helpful comments, for
sharing\ his broad knowledge, and also for a careful reading of the
manuscript, and sincerely thanks J. I. Cirac for enlightening guidance and
encouragement. He is also grateful to Z.-X. Liu, Y. Zhou, X.-G. Wen, T.-K.
Ng, and R. Or\'{u}s for collaborations on related topics, and F. Mezzacapo,
A. E. B. Nielsen, M. Cheng, X.-L. Qi, and A. M\"{u}ller-Hermes for
stimulating discussions. This work has been supported by the EU project
AQUTE.

\onecolumngrid
\appendix
\setcounter{equation}{0}
\newpage

\begin{center}
\textbf{Supplemental material}
\end{center}

\section{Equivalence of the projected BCS state and the infinite MPS}

In this Section, we prove the equivalence of the projected BCS state and the
infinite MPS.

\subsection{Projected BCS state}

Let us first expand the projected BCS state%
\begin{eqnarray*}
|\Psi \rangle &=&P_{\mathrm{G}}\exp \left( \sum_{i<j}\frac{1}{z_{i}-z_{j}}%
\sum_{a=1}^{n}c_{i,a}^{\dagger }c_{j,a}^{\dagger }\right) |0\rangle \\
&=&P_{\mathrm{G}}\prod_{a=1}^{n}\prod_{i<j}\left( 1+\frac{1}{z_{i}-z_{j}}%
c_{i,a}^{\dagger }c_{j,a}^{\dagger }\right) |0\rangle \\
&=&P_{\mathrm{G}}\prod_{a=1}^{n}\left[ \sum_{N_{a}=0\text{ }(N_{a}\text{ even%
})}^{N}\sum_{x_{1}^{(a)}<\cdots <x_{N_{a}}^{(a)}}\mathrm{Pf}_{a}(\frac{1}{%
z_{i}-z_{j}})c_{x_{1}^{(a)},a}^{\dagger }\cdots
c_{x_{N_{a}}^{(a)},a}^{\dagger }\right] |0\rangle \\
&=&P_{\mathrm{G}}\sum_{N_{1},N_{2},\ldots ,N_{n}=0\text{ }(N_{a}\text{ even}%
)}^{N}\sum_{x_{1}^{(1)}<\cdots <x_{N_{1}}^{(1)}}\sum_{x_{2}^{(2)}<\cdots
<x_{N_{2}}^{(2)}}\cdots \sum_{x_{1}^{(n)}<\cdots
<x_{N_{a}}^{(n)}}\prod_{a=1}^{n}\mathrm{Pf}_{a}(\frac{1}{z_{i}-z_{j}}) \\
&&\times (c_{x_{1}^{(1)},1}^{\dagger }\cdots c_{x_{N_{1}}^{(1)},1}^{\dagger
})(c_{x_{1}^{(2)},2}^{\dagger }\cdots c_{x_{N_{2}}^{(2)},2}^{\dagger
})\cdots (c_{x_{1}^{(n)},n}^{\dagger }\cdots c_{x_{N_{n}}^{(n)},n}^{\dagger
})|0\rangle
\end{eqnarray*}%
where $N_{a}$ ($a=1,\ldots ,n$) is the number of $c_{a}^{\dagger }$ fermions
(i.e. $|n^{a}\rangle $ vector in the configuration), $x_{1}^{(a)}<\cdots
<x_{N_{a}}^{(a)}$ are the positions of $c_{a}^{\dagger }$ fermions in the
lattice, and the Pfaffian factor $\mathrm{Pf}_{a}(\frac{1}{z_{i}-z_{j}})$ is
restricted to the positions for the $c_{a}^{\dagger }$ fermions.

The next step is to implement the Gutzwiller projection. Note that the
Gutzwiller projector $P_{\mathrm{G}}$ requires single occupancy. Thus, the
positions of fermions ($x_{1}^{(1)}<\cdots <x_{N_{1}}^{(1)}$, $%
x_{2}^{(2)}<\cdots <x_{N_{2}}^{(2)}$, \ldots\ , $x_{1}^{(n)}<\cdots
<x_{N_{a}}^{(n)}$) must be all different from each other,\textbf{\ }so that
each site has exactly one fermion. As a result, we have $%
\sum_{a=1}^{n}N_{a}=N$. After implementing the Gutzwiller projector, we
obtain%
\begin{eqnarray*}
|\Psi \rangle &=&\sum_{N_{1},N_{2},\ldots ,N_{n}=0\text{ (}%
N_{1}+N_{2}+\cdots +N_{n}=N\text{ and }N_{a}\text{ even)}}^{N}\sum_{\text{%
all allowed }x_{1}^{(a)}<\cdots <x_{N_{a}}^{(a)}}\prod_{a=1}^{n}\mathrm{Pf}%
_{a}(\frac{1}{z_{i}-z_{j}}) \\
&&\times (c_{x_{1}^{(1)},1}^{\dagger }\cdots c_{x_{N_{1}}^{(1)},1}^{\dagger
})(c_{x_{1}^{(2)},2}^{\dagger }\cdots c_{x_{N_{2}}^{(2)},2}^{\dagger
})\cdots (c_{x_{1}^{(n)},n}^{\dagger }\cdots c_{x_{N_{n}}^{(n)},n}^{\dagger
})|0\rangle
\end{eqnarray*}

The final step is to rearrange the positions of fermionic operators so that
they can be identified as a spin state. This rearrangement only results in a
sign, depending on the positions of fermions%
\begin{eqnarray*}
|\Psi \rangle &=&\sum_{N_{1},N_{2},\ldots ,N_{n}=0\text{ (}%
N_{1}+N_{2}+\cdots +N_{n}=N\text{ and }N_{a}\text{ even)}}^{N}\sum_{\text{%
all allowed }x_{1}^{(a)}<\cdots <x_{N_{a}}^{(a)}}\prod_{a=1}^{n}\mathrm{Pf}%
_{a}(\frac{1}{z_{i}-z_{j}}) \\
&&\times \mathrm{sgn}(x_{1}^{(1)},\ldots ,x_{N_{1}}^{(1)},x_{1}^{(2)},\ldots
,x_{N_{2}}^{(2)},\ldots ,x_{1}^{(n)},\ldots
,x_{N_{n}}^{(n)})|x_{1}^{(1)},\ldots ,x_{N_{1}}^{(1)},x_{1}^{(2)},\ldots
,x_{N_{2}}^{(2)},\ldots ,x_{1}^{(n)},\ldots ,x_{N_{n}}^{(n)}\rangle
\end{eqnarray*}%
where $|x_{1}^{(1)},\ldots ,x_{N_{1}}^{(1)},x_{1}^{(2)},\ldots
,x_{N_{2}}^{(2)},\ldots ,x_{1}^{(1)},\ldots ,x_{N_{n}}^{(1)}\rangle $ is a
spin configuration labeled by the positions of the vector $|n^{a}\rangle $\
and $\mathrm{sgn}(x_{1}^{(1)},\ldots ,x_{N_{1}}^{(1)},x_{1}^{(2)},\ldots
,x_{N_{2}}^{(2)},\ldots ,x_{1}^{(n)},\ldots ,x_{N_{n}}^{(n)})$ is the
signature of permutation due to the sign factor coming from fermionic
anticommutation relations.

Thus, the projected BCS wave function can be written as%
\begin{equation}
\Psi (\{x^{(1)}\},\{x^{(2)}\},\ldots \{x^{(n)}\})=\mathrm{sgn}%
(x_{1}^{(1)},\ldots ,x_{N_{1}}^{(1)},x_{1}^{(2)},\ldots
,x_{N_{2}}^{(2)},\ldots ,x_{1}^{(n)},\ldots ,x_{N_{n}}^{(n)})\prod_{a=1}^{n}%
\mathrm{Pf}_{a}(\frac{1}{z_{i}-z_{j}})  \label{eq:ProjectedBCS}
\end{equation}%
where $\{x^{(a)}\}$ is the set of positions satisfying $x_{1}^{(a)}<\cdots
<x_{N_{a}}^{(a)}$ ($N_{a}$ even and $\sum_{a=1}^{n}N_{a}=N$).

\subsection{Infinite MPS}

Let us now consider the infinite MPS%
\begin{equation*}
|\Psi \rangle =\sum_{a_{1},\ldots ,a_{N}=1}^{n}\Psi (a_{1},\ldots
,a_{N})|n^{a_{1}},n^{a_{2}},\ldots ,n^{a_{N}}\rangle
\end{equation*}%
where $\Psi (a_{1},\ldots ,a_{N})$ are given by the chiral correlators of
Majorana fermion fields $\chi ^{a}$ ($a=1,\ldots ,n$)
\begin{equation*}
\Psi (a_{1},\ldots ,a_{N})=\langle \chi ^{a_{1}}(z_{1})\chi
^{a_{2}}(z_{2})\cdots \chi ^{a_{N}}(z_{N})\rangle
\end{equation*}

To evaluate $\Psi (a_{1},\ldots ,a_{N})$, we use the two-point correlator of
Majorana fermions
\begin{equation*}
\langle \chi ^{a}(z)\chi ^{b}(w)\rangle =\frac{\delta _{ab}}{z-w}
\end{equation*}%
The multipoint correlators of Majorana fermions are obtained by Wick's
theorem
\begin{equation*}
\langle \chi ^{a}(z_{1})\chi ^{a}(z_{2})\cdots \chi ^{a}(z_{N_{a}})\rangle
=\left\{
\begin{array}{c}
\mathrm{Pf}_{a}(\frac{1}{z_{i}-z_{j}}) \\
0%
\end{array}%
\right. \left.
\begin{array}{c}
N_{a}\text{ even} \\
N_{a}\text{ odd}%
\end{array}%
\right.
\end{equation*}%
Therefore, to obtain a nonvanishing $\Psi (a_{1},\ldots ,a_{N})$, we must
have even $N$. Additionally, $N_{a}$, the number of vectors $|n^{a}\rangle $
in the spin configuration, must also be even for all $a=1,\ldots ,n$.

To compare with the projected BCS wave function, let us evaluate the
superposition coefficient of the infinite MPS for a spin configuration,
which has $N_{a}$ vector $|n^{a}\rangle $ at positions $x_{1}^{(a)}<\cdots
<x_{N_{a}}^{(a)}$ ($N_{a}$ even and $\sum_{a=1}^{n}N_{a}=N$). Taking into
account the anticommuting nature of Majorana fermion fields, we first pick
up the vectors $|n^{1}\rangle $ in the spin configuration and rewrite the
infinite MPS as
\begin{eqnarray*}
\Psi (\{x^{(1)}\},\{x^{(2)}\},\ldots \{x^{(n)}\}) &=&\mathrm{sgn}%
(x_{1}^{(1)},\ldots ,x_{N_{1}}^{(2)},y_{1},\ldots ,y_{N-L_{1}})\langle \chi
^{a=1}(z_{x_{1}^{(1)}})\chi ^{a=1}(z_{x_{2}^{(1)}})\cdots \chi
^{a=1}(z_{x_{N_{1}}^{(1)}})\rangle \\
&&\times \langle \chi ^{b}(z_{y_{1}})\chi ^{b}(z_{y_{2}})\cdots \chi
^{b}(z_{y_{N-N_{1}}})\rangle \text{ \ \ \ \ \ (}b\neq 1\text{)} \\
&=&\mathrm{sgn}(x_{1}^{(1)},\ldots ,x_{N_{1}}^{(2)},y_{1},\ldots
,y_{N-L_{1}})\mathrm{Pf}_{a=1}(\frac{1}{z_{i}-z_{j}}) \\
&&\times \langle \chi ^{b}(z_{y_{1}})\chi ^{b}(z_{y_{2}})\cdots \chi
^{b}(z_{y_{N-N_{1}}})\rangle
\end{eqnarray*}%
where the positions $y_{1}<\cdots <y_{N-N_{1}}$ correspond to the vectors $%
|n^{b}\rangle $ with $b\neq 1$. The above steps can be repeated from $b=2$
to $n$. In the end, we obtain%
\begin{equation}
\Psi (\{x^{(1)}\},\{x^{(2)}\},\ldots \{x^{(n)}\})=\mathrm{sgn}%
(x_{1}^{(1)},\ldots ,x_{N_{1}}^{(1)},x_{1}^{(2)},\ldots
,x_{N_{2}}^{(2)},\ldots ,x_{1}^{(n)},\ldots ,x_{N_{n}}^{(n)})\prod_{a=1}^{n}%
\mathrm{Pf}_{a}(\frac{1}{z_{i}-z_{j}})  \label{eq:iMPS}
\end{equation}

Comparing with Eq. (\ref{eq:ProjectedBCS}), we conclude that the infinite
MPS and the projected BCS state are equivalent.

\section{Derivation of the parent Hamiltonian}

In this Section, we derive the parent Hamiltonian for the infinite MPS.

\subsection{Brief summary of the SO($n$)$_{1}$ WZW model}

For infinite MPS associated to WZW models, the derivation of the parent
Hamiltonian relies on the existence of null vectors in the representation
spaces of Kac-Moody algebra \cite{Anne-2011}. For SO($n$)$_{1}$ WZW model,
the Kac-Moody algebra is defined by%
\begin{equation}
\lbrack J_{n}^{ab},J_{m}^{cd}]=if^{ab,cd,ef}J_{n+m}^{ef}+n\delta
_{ab,cd}\delta _{n+m,0}\text{ \ \ \ \ \ }n,m\in
%TCIMACRO{\U{2124} }%
%BeginExpansion
\mathbb{Z}
%EndExpansion
\label{eq:KMalgebra}
\end{equation}%
where repeated indices are summed over and the SO($n$) structure constant $%
f^{ab,cd,ef}$ is given by%
\begin{equation*}
f^{ab,cd,ef}=\delta _{ad}\delta _{be}\delta _{cf}+\delta _{bc}\delta
_{ae}\delta _{df}-\delta _{ac}\delta _{be}\delta _{df}-\delta _{bd}\delta
_{ae}\delta _{cf}
\end{equation*}

For odd $n$ ($n\geq 3$), the SO($n$)$_{1}$ WZW model has three primary
fields respectively in singlet (denoted by $I$), vector ($v$) and spinor
representation ($s$), whose conformal weights are $h_{I}=0$, $h_{v}=1/2$ and
$h_{s}=n/16$, respectively. For even $n$ ($n\geq 4$), apart from the primary
fields in singlet and vector representations ($h_{I}=0$ and $h_{v}=1/2$),\
the SO($n$)$_{1}$ WZW model has two primary fields in spinor representations
(denoted by $s_{+}$ and $s_{-}$), whose conformal weights are $%
h_{s_{+}}=h_{s_{-}}=n/16$. The SO($n$)$_{1}$ WZW model has central charge $%
c=n/2$ and can be constructed by combining $n$ Ising models ($c=n\times
\frac{1}{2}$).

For both odd and even $n$, the primary fields in the vector representation
have conformal weight $h_{v}=1/2$ and are naturally interpreted as Majorana
fermions, which are the key ingredients for us to construct the infinite MPS.

For each Majorana fermion $\chi ^{a}$ ($a=1,\ldots ,n$), a primary state $%
|\chi ^{a}\rangle $ can be defined by%
\begin{equation*}
|\chi ^{a}\rangle =\chi ^{a}(0)|0\rangle
\end{equation*}%
where $|0\rangle $ is the vacuum of the WZW model and satisfies $%
J_{n>0}^{ab}|0\rangle =0$. When acting on the Kac-Moody currents, the
primary states satisfy%
\begin{eqnarray}
J_{0}^{ab}|\chi ^{c}\rangle &=&-\sum_{d=1}^{n}(L^{ab})_{cd}|\chi ^{d}\rangle
\notag \\
J_{n}^{ab}|\chi ^{c}\rangle &=&0\text{ \ \ \ \ \ (}n>0\text{)}
\label{eq:PrimaryState}
\end{eqnarray}%
where $L^{ab}$ are given by%
\begin{equation*}
(L^{ab})_{cd}=i(\delta _{ac}\delta _{bd}-\delta _{bc}\delta _{ad})
\end{equation*}%
Note that $L^{ab}$ form a closed SO($n$) algebra%
\begin{eqnarray*}
\lbrack L^{ab},L^{cd}] &=&i(\delta _{ad}L^{bc}+\delta _{bc}L^{ad}-\delta
_{ac}L^{bd}-\delta _{bd}L^{ac}) \\
&=&if^{ab,cd,ef}L^{ef}
\end{eqnarray*}

\subsection{Null vectors and parent Hamiltonian}

To derive the parent Hamiltonian, one has to find the null vectors in the SO(%
$n$)$_{1}$ Kac-Moody algebra. The null vectors are descendant states
satisfying%
\begin{equation*}
J_{n}^{ab}|\phi \rangle =0\text{ \ \ \ \ \ (}n>0\text{)}
\end{equation*}

For our purpose, we look for null vectors with the following form:%
\begin{equation*}
|\phi ^{d}\rangle =\sum_{a<b,c}W_{abc}^{d}J_{-1}^{ab}|\chi ^{c}\rangle
\end{equation*}%
where $W_{abc}^{d}$ are the coefficients that have to be determined. They
satisfy the orthonormal condition%
\begin{equation*}
\sum_{a<b,c}(W_{abc}^{d^{\prime }})^{\ast }W_{abc}^{d}=\delta _{d^{\prime }d}
\end{equation*}

In general, the tensor $W_{abc}^{d}$ corresponds to a Clebsch-Gordan
decomposition. For the SU(2)$_{k}$ WZW model, the SU(2) Clebsch-Gordan
coefficients are known \cite{Anne-2011}. However, we are not aware of a
closed form for the SO($n$) Clebsch-Gordan coefficients. To overcome this
difficulty, let us consider the norm of $|\phi ^{d}\rangle $%
\begin{eqnarray*}
\langle \phi ^{d}|\phi ^{d}\rangle  &=&\sum_{a^{\prime }<b^{\prime
},c^{\prime }}\sum_{a<b,c}(W_{a^{\prime }b^{\prime }c^{\prime }}^{d})^{\ast
}W_{abc}^{d}\langle \chi ^{c^{\prime }}|J_{1}^{a^{\prime }b^{\prime
}}J_{-1}^{ab}|\chi ^{c}\rangle  \\
&=&\sum_{a^{\prime }<b^{\prime },c^{\prime }}\sum_{a<b,c}W_{a^{\prime
}b^{\prime }c^{\prime }}^{d}M_{a^{\prime }b^{\prime }c^{\prime
},abc}W_{abc}^{d} \\
&=&(W^{d})^{\dagger }MW^{d}
\end{eqnarray*}%
where $W^{d}$ is viewed as a column vector and $M$ is a matrix defined by
\begin{equation*}
M_{a^{\prime }b^{\prime }c^{\prime },abc}=\langle \chi ^{c^{\prime
}}|J_{1}^{a^{\prime }b^{\prime }}J_{-1}^{ab}|\chi ^{c}\rangle
\end{equation*}%
If $|\phi ^{d}\rangle $ is a null state, $\langle \phi ^{d}|\phi ^{d}\rangle
=(W^{d})^{\dagger }MW^{d}=0$. Since $M$ comes from the norm of two
descendent states, it is a positive-semidefinite matrix satisfying $%
(W^{d})^{\dagger }MW^{d}\geq 0$. Therefore, identifying the orthonormal
vectors $W^{d}$ that belong to the kernel of $M$ gives us all null states $%
|\phi ^{d}\rangle $. For our SO($n$)$_{1}$ WZW model, we can write down the
explicit form of $M$%
\begin{eqnarray*}
M_{a^{\prime }b^{\prime }c^{\prime },abc} &=&\langle \chi ^{c^{\prime
}}|J_{1}^{a^{\prime }b^{\prime }}J_{-1}^{ab}|\chi ^{c}\rangle  \\
&=&\langle \chi ^{c^{\prime }}|[J_{1}^{a^{\prime }b^{\prime
}},J_{-1}^{ab}]|\chi ^{c}\rangle  \\
&=&\langle \chi ^{c^{\prime }}|\left( i\sum_{ef}f^{a^{\prime }b^{\prime
},ab,ef}J_{0}^{ef}+\delta _{a^{\prime }b^{\prime },ab}\right) |\chi
^{c}\rangle  \\
&=&\langle \chi ^{c^{\prime }}|\left( -i\sum_{ef}f^{a^{\prime }b^{\prime
},ab,ef}\sum_{g}(L^{ef})_{cg}|\chi ^{g}\rangle +\delta _{a^{\prime
}b^{\prime },ab}|\chi ^{c}\rangle \right)  \\
&=&-i\sum_{ef}f^{a^{\prime }b^{\prime },ab,ef}(L^{ef})_{cc^{\prime }}+\delta
_{a^{\prime }b^{\prime },ab}\delta _{c^{\prime },c}
\end{eqnarray*}%
where we used Kac-Moody algebra (\ref{eq:KMalgebra}) and the properties of
the primary state (\ref{eq:PrimaryState}). In this way, the null vectors for
the SO($n$)$_{1}$ WZW model are obtained.

Let us mention that the above approach is a systematic way of finding null
vectors and can be easily generalized to other WZW models. The role of the
positive-semidefinite matrix $M$ is similar to the Gram matrix for defining
the Kac determinant \cite{YellowBook} in conformal field theory.

After obtaining $W_{abc}^{d}$ for all the null vectors, we define the
following $K$ tensor \cite{Anne-2011}%
\begin{equation*}
K_{a^{\prime }b^{\prime },c^{\prime }}^{ab,c}=\sum_{d}(W_{a^{\prime
}b^{\prime }c^{\prime }}^{d})^{\ast }W_{abc}^{d}\text{ \ \ \ \ \ (}a<b\text{
and }a^{\prime }<b^{\prime }\text{)}
\end{equation*}%
Let us write these $K$ tensors as $n\times n$ matrices, $K_{a^{\prime
}b^{\prime },c^{\prime }}^{ab,c}=(K_{a^{\prime }b^{\prime
}}^{ab})_{c,c^{\prime }}$. Then, $K_{a^{\prime }b^{\prime }}^{ab}$ have the
following compact form%
\begin{equation*}
K_{a^{\prime }b^{\prime }}^{ab}=\frac{2}{3}\delta _{ab,a^{\prime }b^{\prime
}}-\frac{n+2}{6(n-1)}if^{ab,a^{\prime }b^{\prime },cd}L^{cd}+\frac{n-4}{%
6(n-1)}(L^{ab}L^{a^{\prime }b^{\prime }}+L^{a^{\prime }b^{\prime }}L^{ab})
\end{equation*}

Following Ref. \cite{Anne-2011}, we define an operator $\Lambda _{i}^{ab}$ ($%
1\leq a<b\leq n$)%
\begin{eqnarray*}
\Lambda _{i}^{ab} &=&\sum_{j=1(\neq
i)}^{N}w_{ij}\sum_{c<d}(K^{(i)})_{cd}^{ab}L_{j}^{cd} \\
&=&\sum_{j=1(\neq i)}^{N}w_{ij}\left[ \frac{2}{3}L_{j}^{ab}-\frac{n+2}{6(n-1)%
}if^{ab,cd,ef}L_{i}^{ef}L_{j}^{cd}+\frac{n-4}{6(n-1)}%
(L_{i}^{ab}L_{i}^{cd}+L_{i}^{cd}L_{i}^{ab})L_{j}^{cd}\right] \\
&=&\sum_{j=1(\neq i)}^{N}w_{ij}\left[ \frac{2}{3}L_{j}^{ab}-\frac{1}{n-1}%
L_{i}^{ab}(\vec{L}_{i}\cdot \vec{L}_{j})+\frac{1}{3}(\vec{L}_{i}\cdot \vec{L}%
_{j})L_{i}^{ab}\right]
\end{eqnarray*}%
where $w_{ij}\equiv (z_{i}+z_{j})/(z_{i}-z_{j})$ and $\vec{L}_{i}\cdot \vec{L%
}_{j}\equiv \sum_{a<b}L_{i}^{ab}L_{j}^{ab}$. These operators annihilate the
infinite MPS, $\Lambda _{i}^{ab}|\Psi \rangle =0$ $\forall i,a,b$. Moreover,
the infinite MPS is an SO($n$) singlet and therefore $\sum_{i}L_{i}^{ab}|%
\Psi \rangle =0$ $\forall a,b$. Then, an SO($n$) symmetric parent
Hamiltonian can be defined by%
\begin{equation*}
H=\sum_{i,a<b}(\Lambda _{i}^{ab})^{\dagger }\Lambda
_{i}^{ab}+J\sum_{a<b}(\sum_{i}L_{i}^{ab})^{2}+E_{0}\text{ \ \ \ \ \ (}J\geq 0%
\text{)}
\end{equation*}%
whose ground state is the infinite MPS with energy $E_{0}$.

In 1D, we use $z_{j}=\exp (i\frac{2\pi }{N}j)$ to ensure translational
invariance.\ Choosing $J=2(N-2)/3$ and $E_{0}=-2(n-1)N(N^{2}-4)/9$, we
arrive at the following explicit form of $H$:%
\begin{equation}
H=-\sum_{i\neq j}w_{ij}^{2}[\frac{n+2}{3}(\vec{L}_{i}\cdot \vec{L}_{j})+%
\frac{n-4}{3(n-1)}(\vec{L}_{i}\cdot \vec{L}_{j})^{2}]-\frac{n-4}{3(n-1)}%
\sum_{i\neq j\neq k}w_{ij}w_{ik}(\vec{L}_{i}\cdot \vec{L}_{j})(\vec{L}%
_{i}\cdot \vec{L}_{k}).  \label{eq:Hamiltonian}
\end{equation}%
To obtain the above form, the following identities are quite useful%
\begin{eqnarray*}
\sum_{a<b}L_{i}^{ab}(\vec{L}_{i}\cdot \vec{L}_{j})L_{i}^{ab} &=&\vec{L}%
_{i}\cdot \vec{L}_{j}\text{ \ \ \ \ \ (}i\neq j\text{)} \\
\sum_{a<b}L_{i}^{ab}(\vec{L}_{i}\cdot \vec{L}_{j})^{2}L_{i}^{ab}
&=&2(n-1)-(n-2)(\vec{L}_{i}\cdot \vec{L}_{j})-(\vec{L}_{i}\cdot \vec{L}%
_{j})^{2}\text{ \ \ \ \ \ (}i\neq j\text{)} \\
\sum_{a<b}L_{i}^{ab}(\vec{L}_{i}\cdot \vec{L}_{j})(\vec{L}_{i}\cdot \vec{L}%
_{k})L_{i}^{ab} &=&2(\vec{L}_{j}\cdot \vec{L}_{k})-(\vec{L}_{i}\cdot \vec{L}%
_{k})(\vec{L}_{i}\cdot \vec{L}_{j})\text{ \ \ \ \ \ (}i\neq j\neq k\text{)}
\end{eqnarray*}

\section{Jastrow and Pfaffian wave functions in Cartan basis}

In this Section, we derive the explicit Jastrow and Pfaffian forms of the
wave functions in Cartan basis.

\subsection{Cartan basis}

Let us first define the Cartan basis.

The SO($n$) algebra is defined by%
\begin{equation*}
\lbrack L^{ab},L^{cd}]=i(\delta _{ad}L^{bc}+\delta _{bc}L^{ad}-\delta
_{ac}L^{bd}-\delta _{bd}L^{ac})
\end{equation*}%
For $n=2l$ and $2l+1$, we can choose at most $l$ (rank of the algebra)\
mutually commuting generators as $L^{12},L^{34},\ldots ,L^{2l-1,2l}$. In the
vector basis, the SO($n$) generators are defined by $L^{ab}=i(|n^{a}\rangle
\langle n^{b}|-|n^{b}\rangle \langle n^{a}|)$ ($1\leq a<b\leq n$).
Diagonalizing the Cartan generators gives us the following Cartan basis:%
\begin{eqnarray*}
|1,0,\ldots \rangle &=&\frac{1}{\sqrt{2}}(|n^{2}\rangle +i|n^{1}\rangle ) \\
|-1,0,\ldots \rangle &=&\frac{1}{\sqrt{2}}(|n^{2}\rangle -i|n^{1}\rangle ) \\
|0,1,0,\ldots \rangle &=&\frac{1}{\sqrt{2}}(|n^{4}\rangle +i|n^{3}\rangle )
\\
|0,-1,0,\ldots \rangle &=&\frac{1}{\sqrt{2}}(|n^{4}\rangle -i|n^{3}\rangle )
\\
&&\vdots \\
|0,0,\ldots ,1\rangle &=&\frac{1}{\sqrt{2}}(|n^{2l}\rangle
+i|n^{2l-1}\rangle ) \\
|0,0,\ldots ,-1\rangle &=&\frac{1}{\sqrt{2}}(|n^{2l}\rangle
-i|n^{2l-1}\rangle )
\end{eqnarray*}

For SO($2l$), the above basis is already complete. For SO($2l+1$), we have
an additional vector $|n^{2l+1}\rangle $, which is annihilated by all Cartan
generators. Thus, we have the following extra vector for SO($2l+1$):
\begin{equation*}
|0,0,\ldots ,0\rangle =|n^{2l+1}\rangle
\end{equation*}

Thus, the Cartan basis for SO($2l$) and SO($2l+1$) can be compactly written
as%
\begin{eqnarray*}
|0,\ldots ,m_{\alpha } &=&\pm 1,\ldots ,0\rangle =\frac{1}{\sqrt{2}}%
(|n^{2\alpha }\rangle \pm i|n^{2\alpha -1}\rangle )\text{ \ \ \ \ \ }(\alpha
=1,\ldots ,l) \\
|0,0,\ldots ,0\rangle &=&|n^{2l+1}\rangle
\end{eqnarray*}%
Note that $m_{\alpha }$ is the eigenvalue of the Cartan generator $%
L^{2\alpha -1,2\alpha }$.

\subsection{Even $n=2l$}

Now we derive the Jastrow and Pfaffian forms of the SO($n$) wave functions
in the Cartan basis. Actually, this goal can be achieved from either the
projected BCS form or the infinite MPS form. In the following we use the
projected BCS form to derive the results.

For SO($2l$), we define the slave-fermion operators in the Cartan basis%
\begin{eqnarray*}
c_{m_{1}=\pm 1}^{\dagger } &=&\frac{1}{\sqrt{2}}(c_{2}^{\dagger }\pm
ic_{1}^{\dagger }) \\
c_{m_{2}=\pm 1}^{\dagger } &=&\frac{1}{\sqrt{2}}(c_{4}^{\dagger }\pm
ic_{3}^{\dagger }) \\
&&\vdots \\
c_{m_{l}=\pm 1}^{\dagger } &=&\frac{1}{\sqrt{2}}(c_{2l}^{\dagger }\pm
ic_{2l-1}^{\dagger })
\end{eqnarray*}%
After changing the basis, the SO($2l$) valence-bond singlet operator in the
projected BCS state is written as%
\begin{equation*}
\sum_{a=1}^{2l}c_{i,a}^{\dagger }c_{j,a}^{\dagger }=\sum_{\alpha
=1}^{l}(c_{i,m_{\alpha }=1}^{\dagger }c_{j,m_{\alpha }=-1}^{\dagger
}+c_{i,m_{\alpha }=-1}^{\dagger }c_{j,m_{\alpha }=1}^{\dagger })
\end{equation*}

Using the above form, the SO($2l$) projected BCS state is rewritten as%
\begin{eqnarray*}
|\Psi _{\mathrm{SO(}2l\mathrm{)}}\rangle &=&P_{\mathrm{G}}\exp \left(
\sum_{i<j}\frac{1}{z_{i}-z_{j}}\sum_{a=1}^{2l}c_{i,a}^{\dagger
}c_{j,a}^{\dagger }\right) |0\rangle \\
&=&P_{\mathrm{G}}\exp \left( \sum_{i\neq j}\frac{1}{z_{i}-z_{j}}\sum_{\alpha
=1}^{l}c_{i,m_{\alpha }=1}^{\dagger }c_{j,m_{\alpha }=-1}^{\dagger }\right)
|0\rangle \\
&=&P_{\mathrm{G}}\prod_{\alpha =1}^{l}\prod_{i\neq j}\left( 1+\frac{1}{%
z_{i}-z_{j}}c_{i,m_{\alpha }=1}^{\dagger }c_{j,m_{\alpha }=-1}^{\dagger
}\right) |0\rangle \\
&=&P_{\mathrm{G}}\prod_{\alpha =1}^{l}\left[ \sum_{N_{\alpha
}=0}^{N/2}\sum_{p_{1}^{(\alpha )}<\cdots <p_{N_{\alpha }}^{(\alpha
)}}\sum_{q_{1}^{(\alpha )}<\cdots <q_{N_{\alpha }}^{(\alpha )}}\det \left(
\frac{1}{z_{i}-z_{j}}\right) _{(p_{1}^{(\alpha )}\cdots p_{N_{\alpha
}}^{(\alpha )}),(q_{1}^{(\alpha )}\cdots q_{N_{\alpha }}^{(\alpha )})}\right.
\\
&&\times \left. c_{p_{1}^{(\alpha )},m_{\alpha }=1}^{\dagger
}c_{q_{1}^{(\alpha )},m_{\alpha }=-1}^{\dagger }\cdots c_{p_{N_{\alpha
}}^{(\alpha )},m_{\alpha }=1}^{\dagger }c_{q_{N_{\alpha }}^{(\alpha
)},m_{\alpha }=-1}^{\dagger }\right] |0\rangle
\end{eqnarray*}%
where $\det \left( \frac{1}{z_{i}-z_{j}}\right) _{(p_{1}^{(\alpha )}\cdots
p_{N_{\alpha }}^{(\alpha )}),(q_{1}^{(\alpha )}\cdots q_{N_{\alpha
}}^{(\alpha )})}$ is the determinant of the $N_{\alpha }\times N_{\alpha }$
Cauchy matrix restricted to the positions of $c_{m_{\alpha }=1}^{\dagger }$
and $c_{m_{\alpha }=-1}^{\dagger }$ fermions. The following useful identity
reduces the Cauchy determinant to a product of Jastrow factors:%
\begin{equation*}
\det \left( \frac{1}{z_{i}-z_{j}}\right) _{(p_{1}^{(\alpha )}\cdots
p_{N_{\alpha }}^{(\alpha )}),(q_{1}^{(\alpha )}\cdots q_{N_{\alpha
}}^{(\alpha )})}=(-1)^{\frac{1}{2}N_{\alpha }(N_{\alpha }-1)}\frac{%
\prod_{1\leq i<j\leq N_{\alpha }}(z_{p_{i}^{(\alpha )}}-z_{p_{j}^{(\alpha
)}})(z_{q_{i}^{(\alpha )}}-z_{q_{j}^{(\alpha )}})}{\prod_{1\leq i,j\leq
N_{\alpha }}(z_{p_{i}^{(\alpha )}}-z_{q_{j}^{(\alpha )}})}
\end{equation*}%
Note that the sign factor in the Cauchy determinant can be absorbed by
rearranging the fermionic operators%
\begin{eqnarray*}
&&c_{p_{1}^{(\alpha )},m_{\alpha }=1}^{\dagger }c_{q_{1}^{(\alpha
)},m_{\alpha }=-1}^{\dagger }c_{p_{2}^{(\alpha )},m_{\alpha }=1}^{\dagger
}c_{q_{2}^{(\alpha )},m_{\alpha }=-1}^{\dagger }\cdots c_{p_{N_{\alpha
}}^{(\alpha )},m_{\alpha }=1}^{\dagger }c_{q_{N_{\alpha }}^{(\alpha
)},m_{\alpha }=-1}^{\dagger } \\
&=&(-1)^{\frac{1}{2}N_{\alpha }(N_{\alpha }-1)}(c_{p_{1}^{(\alpha
)},m_{\alpha }=1}^{\dagger }c_{p_{2}^{(\alpha )},m_{\alpha }=1}^{\dagger
}\cdots c_{p_{N_{\alpha }}^{(\alpha )},m_{\alpha }=1}^{\dagger
})(c_{q_{1}^{(\alpha )},m_{\alpha }=-1}^{\dagger }c_{q_{2}^{(\alpha
)},m_{\alpha }=-1}^{\dagger }\cdots c_{q_{N_{\alpha }}^{(\alpha )},m_{\alpha
}=-1}^{\dagger })
\end{eqnarray*}%
Therefore, we obtain%
\begin{eqnarray*}
|\Psi _{\mathrm{SO(}2l\mathrm{)}}\rangle &=&P_{\mathrm{G}}\prod_{\alpha
=1}^{l}\left[ \sum_{N_{\alpha }=0}^{N/2}\sum_{p_{1}^{(\alpha )}<\cdots
<p_{N_{\alpha }}^{(\alpha )}}\sum_{q_{1}^{(\alpha )}<\cdots <q_{N_{\alpha
}}^{(\alpha )}}\frac{\prod_{1\leq i<j\leq N_{\alpha }}(z_{p_{i}^{(\alpha
)}}-z_{p_{j}^{(\alpha )}})(z_{q_{i}^{(\alpha )}}-z_{q_{j}^{(\alpha )}})}{%
\prod_{1\leq i,j\leq N_{\alpha }}(z_{p_{i}^{(\alpha )}}-z_{q_{j}^{(\alpha
)}})}\right. \\
&&\times \left. (c_{p_{1}^{(\alpha )},m_{\alpha }=1}^{\dagger
}c_{p_{2}^{(\alpha )},m_{\alpha }=1}^{\dagger }\cdots c_{p_{N_{\alpha
}}^{(\alpha )},m_{\alpha }=1}^{\dagger })(c_{q_{1}^{(\alpha )},m_{\alpha
}=-1}^{\dagger }c_{q_{2}^{(\alpha )},m_{\alpha }=-1}^{\dagger }\cdots
c_{q_{N_{\alpha }}^{(\alpha )},m_{\alpha }=-1}^{\dagger })\right] |0\rangle
\end{eqnarray*}%
In the next step, we collect the positions $p_{1}^{(\alpha )}<\cdots
<p_{N_{\alpha }}^{(\alpha )}$ and $q_{1}^{(\alpha )}<\cdots <q_{N_{\alpha
}}^{(\alpha )}$ into a single set $\{x^{(\alpha )}\}$ with $x_{1}^{(\alpha
)}<\cdots <x_{2N_{\alpha }}^{(\alpha )}$. Then, the Jastrow factors can be
written as%
\begin{equation*}
\frac{\prod_{1\leq i<j\leq N_{\alpha }}(z_{p_{i}^{(\alpha
)}}-z_{p_{j}^{(\alpha )}})(z_{q_{i}^{(\alpha )}}-z_{q_{j}^{(\alpha )}})}{%
\prod_{1\leq i,j\leq N_{\alpha }}(z_{p_{i}^{(\alpha )}}-z_{q_{j}^{(\alpha
)}})}\rightarrow \prod_{1\leq i<j\leq 2N_{\alpha }}(z_{x_{i}^{(\alpha
)}}-z_{x_{j}^{(\alpha )}})^{m_{\alpha ,i}m_{\alpha ,j}}
\end{equation*}%
up to a sign factor. However, the sign factor can again be compensated by
rearranging the fermionic operators to the correct order according to $%
x_{1}^{(\alpha )}<\cdots <x_{2N_{\alpha }}^{(\alpha )}$.

The last step is to implement the Gutzwiller projection and switch to the
spin basis. As a result, we obtain the SO($2l$) wave function in Cartan basis%
\begin{equation}
\Psi (\{m\})=\rho _{m}\prod_{\alpha
=1}^{l}\prod_{i<j}(z_{i}-z_{j})^{m_{\alpha ,i}m_{\alpha ,j}}
\label{eq:evenSOn}
\end{equation}%
where $\rho _{m}=\mathrm{sgn}(x_{1}^{(1)},\ldots ,x_{N_{1}}^{(1)},\ldots
,x_{1}^{(l)},\ldots ,x_{N_{l}}^{(l)})$ if $\sum_{i}m_{\alpha ,i}=0$ $\forall
\alpha $ and $\rho _{m}=0$ otherwise.

For $n=4$, one can further simplify Eq. (\ref{eq:evenSOn}) and show that it
is equivalent to a product of two spin-1/2 Haldane-Shastry states, if the
four vectors are interpreted as two spin-1/2 states.

\subsection{Odd $n=2l+1$}

Comparing to SO($2l$), we have an additional slave-fermion operators in the
Cartan basis for SO($2l+1$)%
\begin{equation*}
c_{m=0}^{\dagger }=c_{2l+1}^{\dagger }
\end{equation*}%
Then, the valence-bond operator for SO($2l+1$)\ is expressed as%
\begin{equation*}
\sum_{a=1}^{2l+1}c_{i,a}^{\dagger }c_{j,a}^{\dagger }=c_{i,m=0}^{\dagger
}c_{j,m=0}^{\dagger }+\sum_{\alpha =1}^{l}(c_{i,m_{\alpha }=1}^{\dagger
}c_{j,m_{\alpha }=-1}^{\dagger }+c_{i,m_{\alpha }=-1}^{\dagger
}c_{j,m_{\alpha }=1}^{\dagger })
\end{equation*}

The expansion of the SO($2l+1$) projected BCS state is given by
\begin{eqnarray*}
|\Psi _{\mathrm{SO(}2l+1\mathrm{)}}\rangle &=&P_{\mathrm{G}}\exp \left(
\sum_{i<j}\frac{1}{z_{i}-z_{j}}\sum_{a=1}^{2l+1}c_{i,a}^{\dagger
}c_{j,a}^{\dagger }\right) |0\rangle \\
&=&P_{\mathrm{G}}\exp \left( \sum_{i<j}\frac{1}{z_{i}-z_{j}}%
c_{i,m=0}^{\dagger }c_{j,m=0}^{\dagger }\right) \exp \left( \sum_{i\neq j}%
\frac{1}{z_{i}-z_{j}}\sum_{\alpha =1}^{l}c_{i,m_{\alpha }=1}^{\dagger
}c_{j,m_{\alpha }=-1}^{\dagger }\right) |0\rangle \\
&=&P_{\mathrm{G}}\left[ \sum_{N_{0}=0\text{ (}N_{0}\text{ even)}%
}^{N}\sum_{x_{1}^{(0)}<\cdots x_{N_{0}}^{(0)}}\mathrm{Pf}_{0}\left( \frac{1}{%
z_{i}-z_{j}}\right) c_{x_{1}^{(0)},m=0}^{\dagger }\cdots
c_{x_{N_{0}}^{(0)},m=0}^{\dagger }\right] \\
&&\times \exp \left( \sum_{i\neq j}\frac{1}{z_{i}-z_{j}}\sum_{\alpha
=1}^{l}c_{i,m_{\alpha }=1}^{\dagger }c_{j,m_{\alpha }=-1}^{\dagger }\right)
|0\rangle
\end{eqnarray*}%
where the positions of the extra fermion $c_{m=0}^{\dagger }$ are labeled by
$x_{1}^{(0)}<\cdots <x_{N_{0}}^{(0)}$. The rest of the calculation is
similar to the SO($2l$) case, except for the presence of a Pfaffian factor
due to the extra fermionic mode $c_{m=0}^{\dagger }$. After some algebra, we
obtain the SO($2l+1$) wave function in Cartan basis%
\begin{equation}
\Psi (\{m\})=\rho _{m}\mathrm{Pf}_{0}(\frac{1}{z_{i}-z_{j}})\prod_{\alpha
=1}^{l}\prod_{i<j}(z_{i}-z_{j})^{m_{\alpha ,i}m_{\alpha ,j}}
\label{eq:oddSOn}
\end{equation}%
where $\rho _{m}=\mathrm{sgn}(x_{1}^{(0)}\ldots
x_{N_{0}}^{(0)},x_{1}^{(1)},\ldots ,x_{N_{1}}^{(1)},\ldots
,x_{1}^{(l)},\ldots ,x_{N_{l}}^{(l)})$ if $\sum_{i}m_{\alpha ,i}=0$ $\forall
\alpha $ and $\rho _{m}=0$ otherwise.

For $n=3$, the signature of the permutation in $\rho _{m}$ is reduced to the
Marshall sign in the spin-1 Haldane-Shastry state \cite{Anne-2011}.


\begin{thebibliography}{99}
\bibitem{Laughlin-1983} R. B. Laughlin, Phys. Rev. Lett. \textbf{50}, 1395
(1983).

\bibitem{Bethe-1931} H. Bethe, Z. Phys. \textbf{71}, 205 (1931).

\bibitem{Haldane-Shastry-1988} F. D. M. Haldane, Phys. Rev. Lett. \textbf{60}%
, 635 (1988); B. S. Shastry, \textit{ibid}. \textbf{60}, 639 (1988).

\bibitem{Anderson-1987} P. W. Anderson, Science \textbf{235}, 1196 (1987).

\bibitem{Haldane-1983} F. D. M. Haldane, Phys. Rev. Lett. \textbf{51}, 605
(1983).

\bibitem{Haldane-Rezayi-1985} F. D. M. Haldane and E. H. Rezayi, Phys. Rev.
Lett. \textbf{54}, 237 (1985).

\bibitem{Affleck-1987} I. Affleck, T. Kennedy, E. H. Lieb, and H. Tasaki,
Phys. Rev. Lett. \textbf{59}, 799 (1987).

\bibitem{Ignacio-German-2010} J. I. Cirac and G. Sierra, Phys. Rev. B
\textbf{81}, 104431 (2010).

\bibitem{Anne-2011} A. E. B. Nielsen, J. I. Cirac, and G. Sierra, J. Stat.
Mech. P11014 (2011).

\bibitem{Greiter-2012} R. Thomale, S. Rachel, P. Schmitteckert, and M.
Greiter, Phys. Rev. B \textbf{85}, 195149 (2012); M. Greiter, J. Low Temp.
Phys. \textbf{126}, 1029 (2002).

\bibitem{Paredes-2012} B. Paredes, Phys. Rev. B \textbf{85}, 195150 (2012).

\bibitem{ZXLiu-2012} Z.-X. Liu, Y. Zhou, H.-H. Tu, X.-G. Wen, and T.-K. Ng,
Phys. Rev. B \textbf{85}, 195144 (2012).

\bibitem{Takhtajan-Babujian-1982} L. A. Takhtajan, Phys. Lett. A \textbf{87}%
, 479 (1982); H. M. Babujian, \textit{ibid}. \textbf{90}, 479 (1982).

\bibitem{Affleck-1986} I. Affleck, Phys. Rev. Lett. \textbf{56}, 746 (1986).

\bibitem{YellowBook} P. Di Francesco, P. Mathieu, and D. S\'{e}n\'{e}chal,
\textit{Conformal Field Theory} (Springer, New York, 1997).

\bibitem{Ardonne-German-2010} E. Ardonne and G. Sierra, J. Phys. A \textbf{43%
}, 505402 (2010).

\bibitem{SuppleMatSOn} See Supplemental Material for the proof of the
equivalence of projected BCS state and infinite MPS, derivation of the
parent Hamiltonian, and derivation of the Jastrow and Pfaffian wave
functions in the Cartan basis.

\bibitem{Georgi-1999} H. Georgi, \textit{Lie Algebras in Particle Physics}
(Perseus Books, Reading, MA, 1999).

\bibitem{Cardy-Calabrese-2004} C. Holzhey, F. Larsen, and F. Wilczek, Nucl.
Phys. B \textbf{424}, 443 (1994); G. Vidal, J. I. Latorre, E. Rico, and A.
Kitaev, Phys. Rev. Lett. \textbf{90}, 227902 (2003); P. Calabrese and J.
Cardy, J. Stat. Mech. P06002 (2004).

\bibitem{Horsch-Kaplan-1983} P. Horsch and T. A. Kaplan, J. Phys. C \textbf{%
16}, L1203 (1983).

\bibitem{Tu-2011} H.-H. Tu and R. Or\'us, Phys. Rev. Lett. \textbf{107},
077204 (2011).

\bibitem{Hastings-2010} M. B. Hastings, I. Gonz\'{a}lez, A. B. Kallin, and
R. G. Melko, Phys. Rev. Lett. \textbf{104}, 157201 (2010).

\bibitem{YZhang-2011} Y. Zhang, T. Grover, and A. Vishwanath, Phys. Rev.
Lett. \textbf{107}, 067202 (2011).

\bibitem{Fabio-2012} F. Mezzacapo, Phys. Rev. B \textbf{86}, 045115 (2012).

\bibitem{Kawakami-1992} N. Kawakami, Phys. Rev. B \textbf{46}, 1005 (1992);
\textbf{46}, 3191 (1992).

\bibitem{Ha-Haldane-1992} Z. N. C. Ha and F. D. M. Haldane, Phys. Rev. B.
\textbf{46}, 9359 (1992).

\bibitem{Witten-1984} E. Witten, Commun. Math. Phys. \textbf{92}, 455 (1984).

\bibitem{Gogolin-1998} A. O. Gogolin, A. A. Nersesyan, and A. M. Tsvelik,
\textit{Bosonization and Strongly Correlated Systems} (Cambridge University
Press, Cambridge, England, 1998).

\bibitem{Reshetikhin-1985} N. Y. Reshetikhin, Theor. Math. Phys. \textbf{63}%
, 555 (1985).

\bibitem{Tsvelik-Nersesyan-Schoutens-Lecheminant} A. M. Tsvelik, Phys. Rev.
B \textbf{42}, 10499 (1990); A. A. Nersesyan and A. M. Tsvelik, Phys. Rev.
Lett. \textbf{78}, 3939 (1997); P. Bouwknegt and K. Schoutens, Phys. Rev.
Lett. \textbf{82}, 2757 (1999); F. Alet, S. Capponi, H. Nonne, P.
Lecheminant, and I. P. McCulloch, Phys. Rev. B \textbf{83}, 060407(R) (2011).

\bibitem{Tu-2008} H.-H. Tu, G.-M. Zhang, and T. Xiang, Phys. Rev. B \textbf{%
78}, 094404 (2008); J. Phys. A \textbf{41}, 415201 (2008); H.-H. Tu, G.-M.
Zhang, T. Xiang, Z.-X. Liu, and T.-K. Ng, Phys. Rev. B \textbf{80}, 014401
(2009).

\bibitem{Kolezhuk-Scalapino-Affleck} A. K. Kolezhuk and H. J. Mikeska, Phys.
Rev. Lett. \textbf{80}, 2709 (1998); D. Scalapino, S. C. Zhang, and W.
Hanke, Phys. Rev. B \textbf{58}, 443 (1998); I. Affleck, D. P. Arovas, J. B.
Marston, and D. A. Rabson, Nucl. Phys. B \textbf{366}, 467 (1991).

\bibitem{Kitaev-2001} A. Kitaev, Phys. Usp. \textbf{44}, 131 (2001).

\bibitem{Moore-Read-1991} G. Moore and N. Read, Nucl. Phys. B \textbf{360},
362 (1991).

\bibitem{Anne-2012} A. E. B. Nielsen, J. I. Cirac, and G. Sierra, Phys. Rev.
Lett. \textbf{108}, 257206 (2012).

\bibitem{XGWen-1989} X.-G. Wen, F. Wilczek, and A. Zee, Phys. Rev. B \textbf{%
39}, 11413 (1989).

\bibitem{Read-Green-2000} N. Read and D. Green, Phys. Rev. B \textbf{61},
10267 (2000).

\bibitem{Kitaev-2006} A. Kitaev, Ann. Phys. (NY) \textbf{321}, 2 (2006).

\bibitem{Greiter-2009} M. Greiter and R. Thomale, Phys. Rev. Lett. \textbf{%
102}, 207203 (2009).

\bibitem{Schoutens-1999} P. Bouwknegt and K. Schoutens, Nucl. Phys. B
\textbf{547}, 501 (1999).

\bibitem{Sixteenfoldway} We thank Xiao-Liang Qi and Meng Cheng for sharing
their insights on the 16-fold way of anyonic properties and for
explaining Kitaev's results in Ref. \cite{Kitaev-2006}.

\bibitem{TKNg-PALee-2010} Z.-X. Liu, Y. Zhou, and T.-K. Ng, Phys. Rev. B
\textbf{81}, 224417 (2010); \textbf{82}, 144422 (2010); M. Serbyn, T.
Senthil, and P. A. Lee, Phys. Rev. B \textbf{84}, 180403(R) (2011); S.
Bieri, M. Serbyn, T. Senthil, and P. A. Lee, Phys. Rev. B \textbf{86},
224409 (2012).
\end{thebibliography}
\end{document}